\documentclass[aps,pre,twocolumn,showpacs,groupedaddress,amssymb]{revtex4}

\usepackage{graphicx}
\usepackage{amsmath}
\usepackage{bm}
\usepackage{color}

\begin{document}

\title{Channeling of particles and associated anomalous transport in a 2D complex plasma crystal}
\author{Cheng-Ran Du}
\email{chengran.du@mpe.mpg.de}
\author{Vladimir Nosenko}
\author{Sergey Zhdanov}
\author{Hubertus M. Thomas}
\author{Gregor E. Morfill}
\affiliation{Max-Planck-Institut f\"{u}r extraterrestrische Physik, D-85741 Garching, Germany}

\date{\today}
\begin{abstract}
Implications of recently discovered effect of channeling of upstream extra particles for transport phenomena in a two-dimensional plasma crystal are discussed. Upstream particles levitated above the lattice layer and tended to move between the rows of lattice particles. An example of heat transport is considered, where upstream particles act as moving heat sources, which may lead to anomalous heat transport. The average channeling length observed was $15-20$ interparticle distances. New features of the channeling process are also reported.
\end{abstract}


\maketitle


Complex, or dusty plasmas exist in different environments and take various forms \cite{Ivlev:2012book,VladimirovBook}. One interesting variety is a two-dimensional (2D) plasma crystal \cite{Thomas:1996,Nosenko:2009,Melzer:2013,Nosenko:2007, Nosenko:2011PRL,Nosenko:2012, Nunomura:2002,Piel:02,Couedel:2010,Nosenko:2004,Gavrikov:2005,Hartmann:2011,Nunomura:2005a,Nosenko:2008,Nunomura:2006,Liu:2008}. It is a single-layer suspension of micron-size solid particles in a weakly ionized gas. Due to the high thermal speed of plasma electrons, 
microparticles become negatively charged and interact with each other via a screened Coulomb (Yukawa) pair potential. Under certain conditions, these charged particles can self-organize in a triangular lattice with hexagonal symmetry, forming a 2D plasma crystal. Individual particle motion can be easily recorded in real time using video microscopy. This makes 2D complex plasmas excellent model systems where dynamics can be studied at the level of individual particles which can be regarded as proxy ``atoms.''

Various aspects of 2D complex plasmas that were recently studied in experiments include phase transitions \cite{Thomas:1996,Nosenko:2009,Melzer:2013}, dynamics of defects \cite{Nosenko:2007,Nosenko:2011PRL}, microstructure \cite{Nosenko:2012}, waves \cite{Nunomura:2002,Piel:02}, mode coupling instability \cite{Couedel:2010}, and transport phenomena: momentum \cite{Nosenko:2004,Gavrikov:2005,Hartmann:2011} and heat \cite{Nunomura:2005a,Nosenko:2008} transport, self-diffusion \cite{Nunomura:2006}, superdiffusion \cite{Liu:2008}. Systems other than complex plasmas were also reported to allow superdiffusion: supercooled liquids \cite{Furrukawa:2012}, solid surfaces \cite{Sancho:2004}, granular media \cite{Ganti:2010}.

It was recently discovered that single-layer plasma crystals sometimes include extra particles levitating slightly above the main layer \cite{Du:2012}. Such particles were called upstream particles since they are upstream of the flow of ions in the plasma sheath, with respect to the lattice particles. These extra particles move about and create disturbances in the lattice layer by both Coulomb repulsion and the effective attraction due to the ion wake effect. If a particle moves faster than the sound speed of the plasma crystal, it creates a Mach cone behind it. Compared to other disturbance sources such as a moving extra particle beneath the lattice layer \cite{Samsonov:1999PRL} or a laser beam \cite{Nosenko:2003}, the disturbance is rather weak and thus hard to notice. In certain conditions, the upstream particles tend to travel between the rows of lattice particles, analogous to channeling of energetic particles in regular crystals \cite{Fliller:2006}.

\begin{figure}[!bh]
\includegraphics[width=0.9\columnwidth]{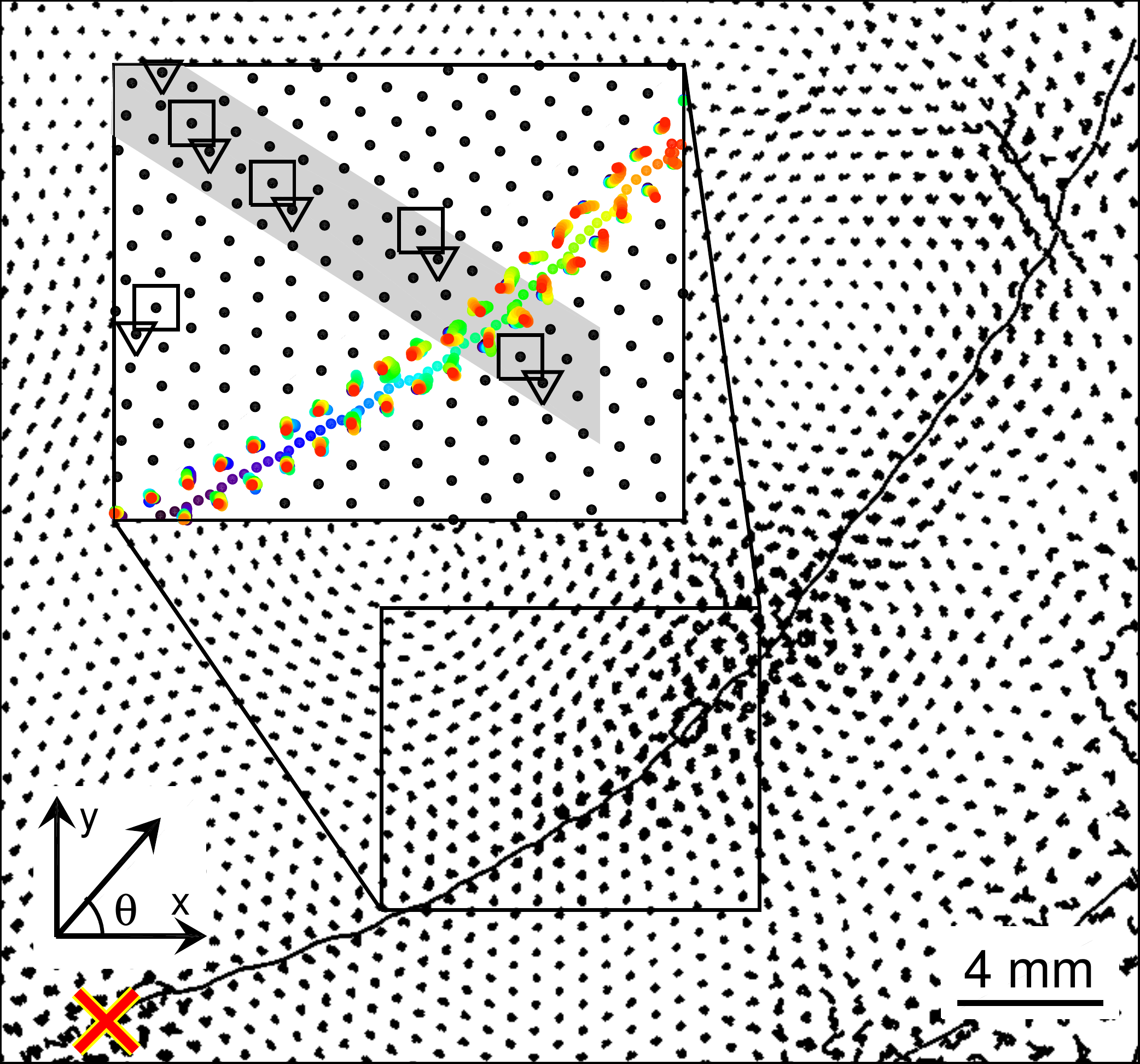}
\caption{(Color) Channeling of an upstream extra particle in a 2D plasma crystal (a movie is also available \cite{movie001}). The figure is an overlay of a sequence of thresholded images recorded by a top-view video camera. The upstream particle trajectory is the curve across the figure starting at the red cross mark.
The inset is a zoom-in of the area where the upstream particle encounters a chain of dislocations. Black dots represent the lattice particles in a snapshot taken $1.5$~s before the upstream particle came inside the field of view. Colored dots represent the ``wall'' particles and the upstream particle positions in a $0.7$-s time sequence coded from blue to red. The chain of dislocations is marked by a gray stripe, the 7-fold and 5-fold cells are indicated by squares and triangles, respectively.
\label{figure1}}
\end{figure}

In this paper, we study channeling of upstream extra particles in a 2D plasma crystal in more detail and show its implications for experiments on transport phenomena.


The experiments were performed in a modified Gaseous Electronics Conference (GEC) rf reference cell \cite{Nosenko:2012}. Argon plasma was sustained using a capacitively coupled rf discharge at $13.56$~MHz. The input power was set at $20$~W. We used monodisperse polystyrene (PS) particles to create a 2D plasma crystal suspended above the bottom rf electrode. The particles have a diameter of $11.36\pm0.12$~$\mu$m and mass density of $1.05$~g/cm$^{3}$. The gas pressure was maintained at $0.65$~Pa, corresponding to the gas friction rate $\gamma\simeq0.91$~s$^{-1}$ \cite{Liu:2003}. The lattice layer was illuminated by a horizontal laser sheet shining through a side window of the chamber. A high-resolution video camera (Photron FASTCAM~1024~PCI) was mounted above the chamber, capturing a top view with a size of $42.7\times42.7$~mm$^2$. The recording rate was set at $250$~frames per second.



We prepared 2D plasma crystals using a standard technique \cite{Nosenko:2011PRL}. After the particles were injected into plasma, they formed a single-layer suspension; in addition, some heavier particles always levitated beneath the main layer \cite{Samsonov:1999PRL,Schweigert:2002}. These were mainly agglomerations of two \cite{Chaudhuri:2012} or more particles \cite{Du:2012PoP}. We removed these particles from the suspension by dropping them on the lower rf electrode while reducing the discharge power. However, even after all heavier particles were removed, we sometimes observed moving disturbances in the lattice. These were caused by the upstream extra particles that levitated {\it above} the main layer as was recently discovered in Ref.~\cite{Du:2012}. In the present paper, we study in more detail the dynamics of upstream extra particles and their effect on the transport phenomena in the main layer. To maximize the rate of occurrence of upstream extra particles, we used polystyrene (PS) particles in this study. The reason why plasma crystals composed of PS particles contained upstream extra particles more often than those composed of melamine formaldehyde (MF) particles is not clear.

An upstream extra particle usually moved about in a horizontal plane approximately $0.25\Delta$ higher than the main layer, where $\Delta$ is the in-plane interparticle distance. Since a moving particle must overcome the ambient gas friction, there must be a driving force acting on it; the origin of this force is not clear. Possible candidates are the wake-field interaction \cite{Schweigert:2002} and photophoretic force from the illuminating laser \cite{Nosenko:2010PoP}. From a different perspective, upstream particles can be regarded as self-propelled (active) particles \cite{Buttinoni:2013}. They may be smaller particles from within the natural size distribution or particles with some defects on their surface. The rather small vertical separation between the main layer and the upstream particles allowed us to illuminate and observe them simultaneously by properly adjusting the height of the illuminating laser sheet.

\begin{figure}
\includegraphics[width=\columnwidth]{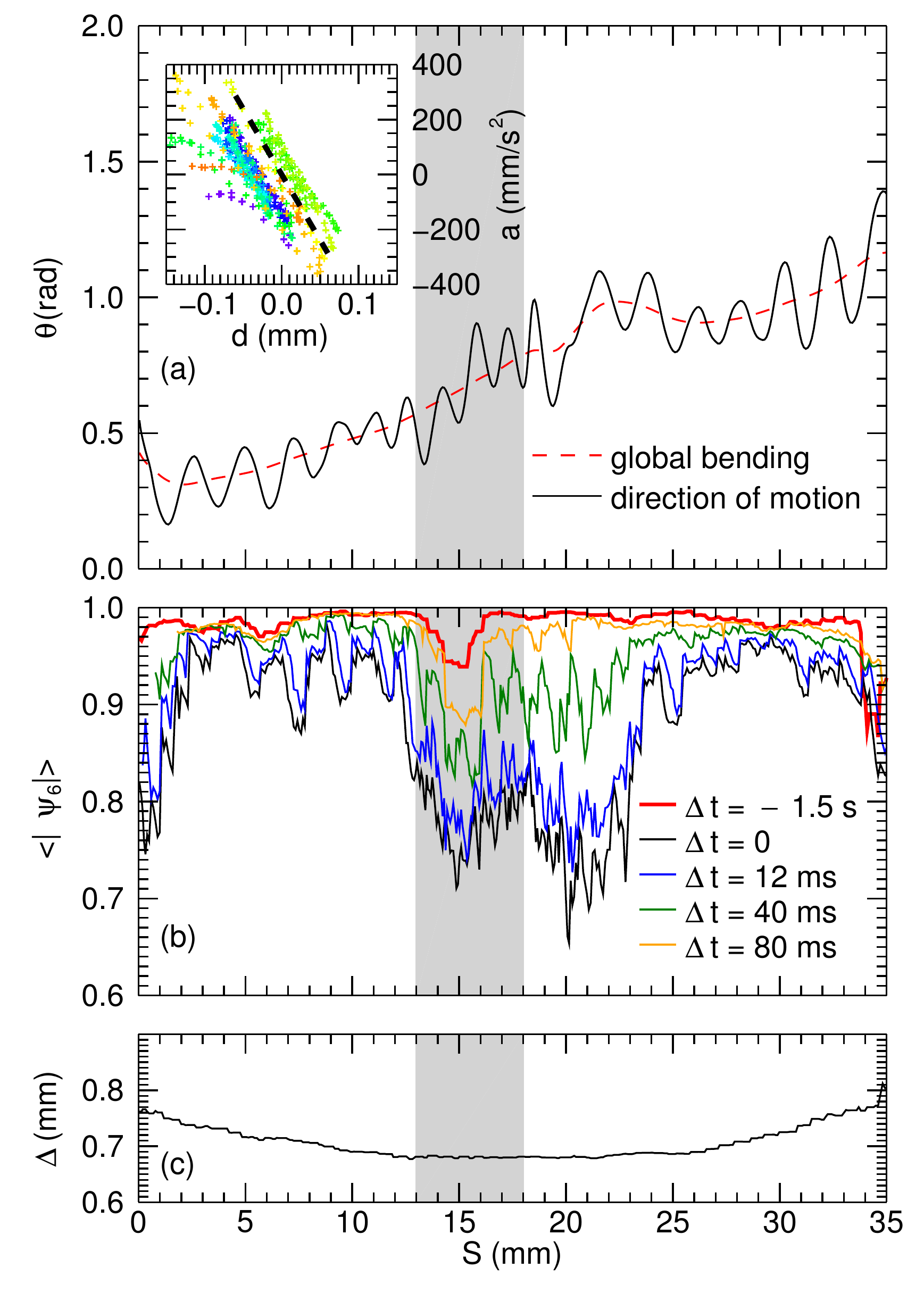}
\caption{(Color) Analysis of the channeling event shown in Fig.~\ref{figure1}. The $S$-axis represents a natural coordinate system along the upstream particle trajectory, where $S=0$ is marked by a red cross in Fig.~\ref{figure1}. Panel (a) shows the instantaneous direction of motion $\theta$ (black) and the global bending angle (red) of the upstream particle. The inset shows the particle transverse acceleration $a$ as a function of its deviation $d$ from the central line of the channel. Color-coding from blue to orange represents timing during $1.85$~s. Panel (b) presents the local bond-orientational order parameter $\langle|\psi_6|\rangle$ (see text) of the lattice at different delay times ${\Delta}t$. Panel (c) shows the local interparticle separation. The gray stripes correspond to the chain of dislocations marked in Fig.~\ref{figure1}.
\label{figure2}}
\end{figure}

A particularly long trajectory of an upstream particle is shown in Fig.~\ref{figure1}. By tracking its motion and analyzing its interaction with the lattice particles, we observed several new features of the channeling process. First, the channeling path is not necessarily straight. As the upstream particle follows the channel formed by the lattice particles, it is constantly redirected and the resulting trajectory may be curved as in Fig.~\ref{figure1}. Here, the total angle of deflection is about $50^{\circ}$, see Fig.~\ref{figure2}(a). The rows of lattice particles are curved due to the presence of defects in the lattice, e.g. a chain of dislocations marked by a grey stripe in Fig.~\ref{figure1} \footnote{At the periphery of the crystal, shell structure \cite{Dubin:1989} may also account for the curved channels.}. This situation is analogous to the channeling effect in regular crystals. It is well known that bent crystals are used to redirect energetic particle beams \cite{Fliller:2006}. Despite the totally different energy, space, and time scales involved, our experiment presents the first direct observation of the dynamics of such process.

Second, we found that the upstream particle scattering angle is determined by two competing factors: number density and regularity of the lattice. As an upstream particle travels in the channel, it bounces on the channel walls and its trajectory has a zig-zag pattern. The scattering angle is related to the lattice particle number density. As shown in Figs.~\ref{figure2}(a),(c), the magnitude of scattering is relatively large at two ends where the local lattice density is relatively low. The scattering angle decreases as particle moves into the central region of the particle suspension, where the particle number density is higher. However, when the upstream particle encounters lattice defects (e.g., a chain of dislocations), the scattering angle increases, see Fig.~\ref{figure2}(a). The channeling process may continue, as in Fig.~\ref{figure2}(a), or it may be interrupted, if the lattice distortion is too large. In Ref.~\cite{Du:2012}, the upstream particle ``jumped'' out of the lattice layer upon encountering a chain of dislocation. Depending on the local structure, this particle either found a new channel and started channeling again or kept jumping until it left the particle suspension.

Third, we gained insight into the channel internal structure by analyzing the particle transverse acceleration $a$
as a function of its deviation $d$ from the central line of the channel, see inset in Fig.~\ref{figure2}(a). The slope here is proportional to the square of the transverse oscillation frequency of channeling according to Hooke's law. The dashed line corresponds to the oscillation frequency of $11$~Hz obtained from the Fourier-transformation analysis of the transverse velocity. The double-strand shape of the $a(d)$ plot suggests that the particle actually moves along one of two lines in the channel closer to its walls and not along the central line.

\begin{figure}
\includegraphics[bb=60 20 800 350, width=\columnwidth]{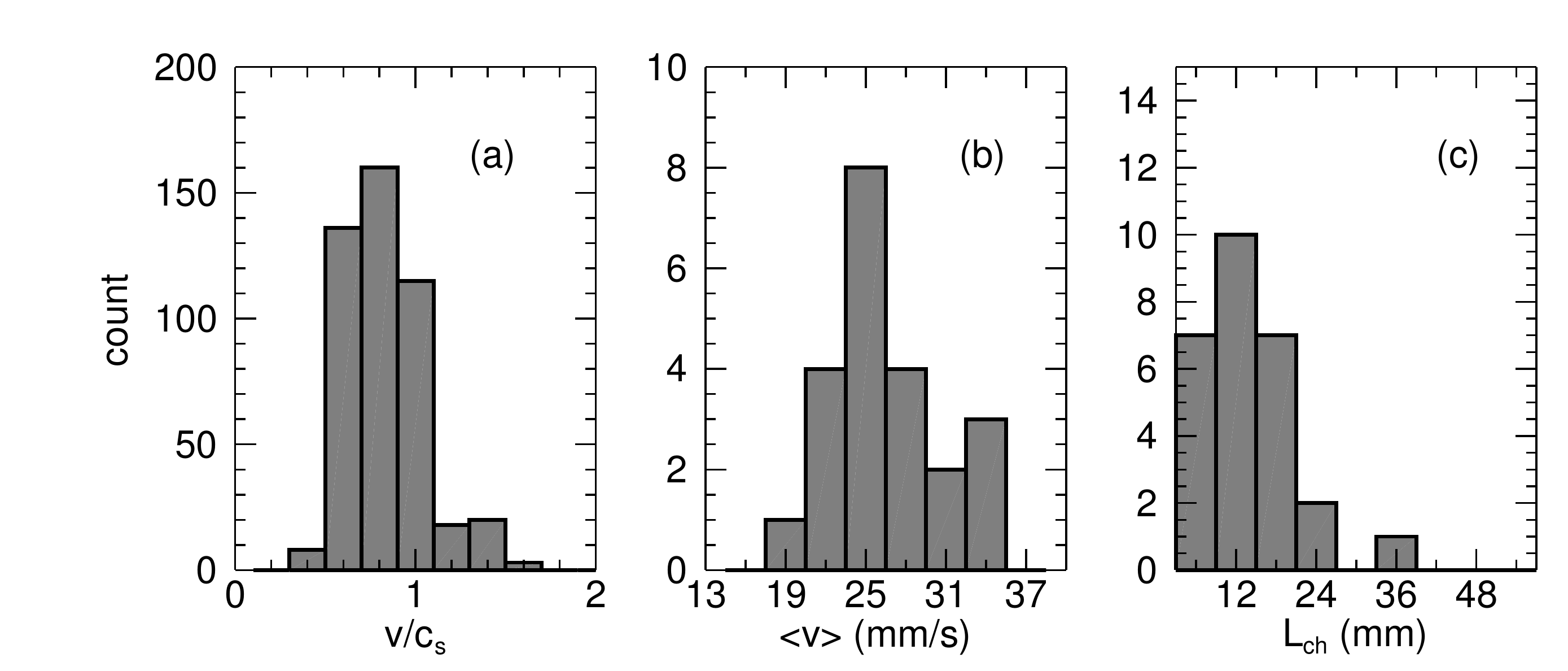}
\caption{Velocity and trajectory length distributions for the observed channeling events. Panel (a) is for the ratio of instantaneous velocity of upstream particle to the local sound speed $C_s$ for the event shown in Fig.~\ref{figure1}. Panels (b) and (c) are respectively for the mean velocity $\langle v\rangle$ and trajectory length $L_{\rm ch}$ of the channeling events from a series of experiment realizations.
\label{figure3}}
\end{figure}

We observed channeling upstream particles in several experiments. In all cases, their speed was comparable to the sound speed of compressional waves $C_s$. In Fig.~\ref{figure3}, we summarize the velocity and trajectory length distributions for such channeling events from a series of experiments with the same conditions. The local sound speed is given by the relation $C_s=2.44\sqrt{{Q^2\lambda_D}/m{\Delta}^2}$, where $\Delta$ is the interparticle separation in the respective lattice location, $Q$ is the particle charge, $\lambda_D$ is screening length, and $m$ is the particle mass \cite{Zhdanov:2002}.

The average channeling length in different experiments was $\langle L_{\rm ch} \rangle \approx 12$~mm, see Fig.~\ref{figure3}(c). Very long trajectories like that in Fig.~\ref{figure1} were rare. Note, however, that $\langle L_{\rm ch} \rangle \gg\Delta$ (the latter was $0.65-0.75$~mm). This may present a problem for studying transport phenomena in this system. For example, if one intends to study diffusion, the existence of such long particle trajectories apparently contradicts the basic assumption that diffusion results from the random walk of particles.

The diffusion coefficient $D$ is usually calculated by measuring the mean squared displacement (MSD) of particles in an ensemble \cite{Ohta:2000}:
\begin{equation}
\label{diffusion}
D=\lim_{t \to \infty} (2nt)^{-1} \langle |{\bf r}_i(t)-{\bf r}_i(0)|^2 \rangle,
\end{equation}
where ${\bf r}_i(t)$ is the position of the $i$th particle and $n$ is the space dimensionality ($n=2$ in our case). If upstream particles are accidentally recorded in an experiment and enter the calculation of MSD in Eq.~(\ref{diffusion}) along with the regular particles, their high velocity (on the order of $C_s$) will lead to a spurious rise in the diffusion coefficient. In fact, in this case MSD will not even have a $\propto t$ asymptote at large times, since the displacement of an upstream particle is (roughly) proportional to time. Therefore, an accurate measurement of diffusion coefficient under these circumstances is problematic.

Furthermore, a moving upstream particle interacts with the lattice particles, particularly those in the channel walls, and transfers some of its momentum and energy to the surrounding particles. The momentum and energy will then spread further in the particle suspension. This may affect any measurements of the transport properties.

To quantify the disturbance created by an upstream particle in the lattice, we use the bond-orientational order parameter defined at each site as $\psi_6=n^{-1}\Sigma_{j=1}^{n}\exp(6i\Theta_j)$, where $\Theta_j$ are bond orientation angles for $n$ nearest neighbors \cite{Grier:1994}. In a perfect hexagonal structure, $|\psi_6|=1$. The smaller $|\psi_6|<1$, the more disordered the structure is, e.g. $|\psi_6|\approx 0.35$ in a dislocation core \cite{Nosenko:2007}. Here, we average $|\psi_6|$ within a small circular area with a radius of $1.2$~mm to quantify the local degree of order in the lattice along the trajectory of upstream particle, see Fig.~\ref{figure2}(b). ${\Delta}t$ is the delay time of the measurement, where negative and positive values correspond respectively to the time before and after the upstream particle reaches the corresponding site along the $S$-axis.

As can be seen in Fig.~\ref{figure2}(b), a dip in $\langle|\psi_6|\rangle$ at $S\approx15$~mm corresponds to a chain of dislocations that was present in the crystal even before the arrival of the upstream particle. As the upstream particle travels in the channel between $S\approx3$~mm and $\approx13$~mm, the local lattice structure is slightly disturbed with a decrease of $\langle|\psi_6|\rangle$ from $0.99$ to $0.93$. However, when the upstream particle encounters the chain of defects, it starts to move more irregularly and the scattering angle increases. Though it is still confined in the channel, the disturbance in the lattice becomes significant and $\langle|\psi_6|\rangle$ drops from $0.93$ to $0.74$ (black and blue curves). It takes about $80$~ms for the system to restore initial order; the chain of defects is preserved thereafter, as shown by the yellow curve.

\begin{figure}
\includegraphics[width=\columnwidth]{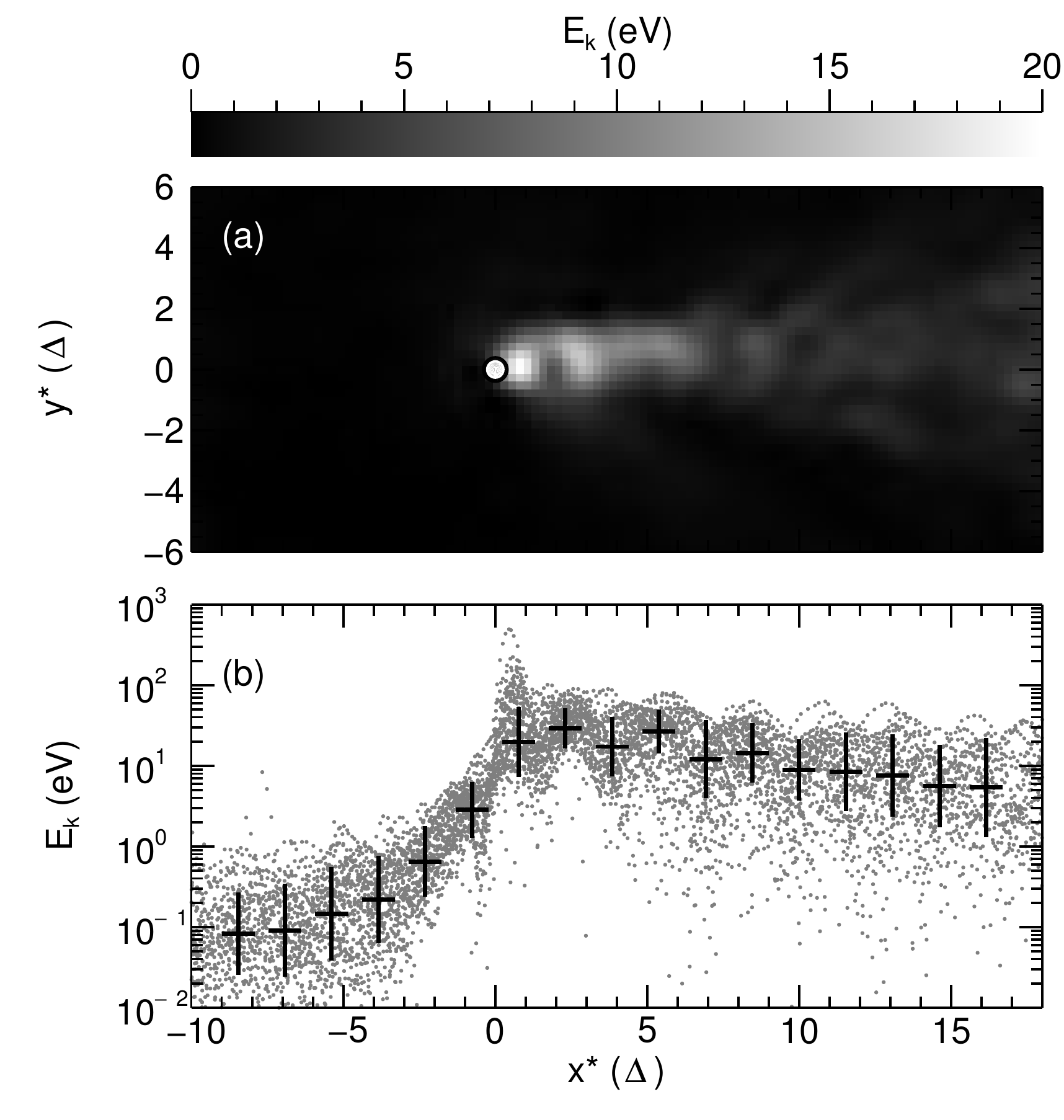}
\caption{Upstream particle as a moving heat source. Panel (a) shows a map of in-plane kinetic energy of the lattice particles averaged from data for 30 consecutive video frames. The upstream particle is marked by a circle located at (0,0) and is moving from right to left. Here $x^*$ and $y^*$ represent the longitudinal and the transverse axes along the trajectory. Panel (b) shows the in-plane kinetic energy $E_k$ of the wall particles. The black plus signs are the means of a gauss-fit of the data along the evolution and error bars represent the standard deviation.
\label{figure4}}
\end{figure}

As an example of the effect of upstream particles on transport phenomena, we consider heat transport in the lattice. As discussed in the previous paragraph, a moving upstream particle disturbs the lattice and generates local disorder. During this process, its kinetic energy is transferred to the lattice particles and therefore it acts as a moving heat source. The transferred heat (in-plane particle kinetic energy) is concentrated on the channel walls for about $5$ interparticle spacings behind the upstream particle, see Fig.~\ref{figure4}. Then the heat starts to spread out transversely. The peak value is located slightly behind the upstream particle and can on average reach a few tens of eV.

The usual analysis of heat transport (based on the idea of heat diffusion) is clearly not applicable in this situation. The moving heat source can be properly taken into account in the following way. If we only concentrate on the kinetic energy of the particles in the walls, namely in the longitudinal direction, we can simplify the heat transport model to a quasi-one-dimensional model \cite{Williams:2012} with a thermal diffusivity $\chi = 2 \gamma L^2 + v L$, where $L$ is the inhomogeneity (gradient) length. The RHS of the equation contains two terms, the first one is associated with conduction and the second with convection. Solving the equation, we obtain $L_{1,2} = \sqrt{(\frac{v}{4\gamma})^2+\frac{\chi}{2\gamma}} \mp \frac{v}{4\gamma}$, corresponding to the inhomogeneity length in front of ($L_1$) and behind ($L_2$) the moving heat source, respectively. The experimental kinetic energy profile can be well fitted by an exponential function $E_k \propto \exp(-|x|/L_{f,b})$. For the inhomogeneity length in front of the upstream particle, $L_f=0.75\pm0.04$~mm. Since the upstream particle travels with a high velocity ($v\approx22$~mm/s), we can simplify $L_1$ to $\chi/v$ and thus from $L_1=L_f$ the thermal diffusivity $\chi=16$~mm$^2$/s. This value is between two previous measurements of $30$~mm$^2$/s \cite{Nunomura:2005a} and $9$~mm$^2$/s \cite{Nosenko:2008} reported in solid and melted 2D complex plasmas, respectively. As to the inhomogeneity length behind the upstream particle, we simplify $L_2$ to $v/2\gamma$, which turns out to be $12$~mm and is associated with the relaxation due to gas friction. However, by fitting the kinetic energy profile behind the upstream particle, we measured $L_b$ to be $6.4\pm0.2$~mm, which is about half of the value $L_2$ predicted by the model. This can be explained by the fact that the heat transport behind the upstream particle has both longitudinal and transverse component. The latter is mainly associated with conduction and thus can be estimated as $L\approx\sqrt{\chi/2\gamma}\approx3$~mm, using the thermal diffusivity obtained earlier. Apart from the in-plane motion of particles, their vertical motion also has a certain contribution to the heat transport. Therefore, our simple one-dimensional model has limited applicability in this case. For other experimental runs we observed that $L_f$ ranges from $0.4$ to $0.8$~mm and $L_b$ ranges from $6$ to $7.1$~mm for the same experimental conditions.


To summarize, we reported new features of the recently discovered effect of upstream extra particle channeling in 2D complex plasma crystals. The implications of particle channeling for transport phenomena were discussed using heat transfer as an example.

So far, we did not find any reliable way of removing the upstream extra particles from the suspension. Since they may pose problems in delicate experiments, it is recommended that the particle suspension be crystalized and carefully checked for the presence of upstream particles. Depending on the particle type and size, the gap between the upstream particles and the main layer may be large and they might not be illuminated by the laser sheet. Therefore, scanning the illumination laser above the lattice layer is recommended.

An important open question is whether particle channeling is also possible in liquid complex plasmas. New dedicated experiments will be necessary to answer this question.


This work was supported by the European Research Council under the European Union's Seventh Framework Programme (FP7/2007-2013) / ERC Grant agreement.

\bibliography{Transport}

\end{document}